\documentclass[twocolumn,english,prl,amsmath,amsfonts,twocolumn]{revtex4-1}
\usepackage[T1]{fontenc}
\usepackage[utf8]{inputenc}
\usepackage{array}
\usepackage{rotating}
\usepackage{bm}
\usepackage{multirow}
\usepackage{graphicx}

\makeatletter

\providecommand{\tabularnewline}{\\}

 \@ifundefined{textcolor}{}
 {%
  \definecolor{BLACK}{gray}{0}
  \definecolor{WHITE}{gray}{1}
  \definecolor{RED}{rgb}{1,0,0}
  \definecolor{GREEN}{rgb}{0,1,0}
  \definecolor{BLUE}{rgb}{0,0,1}
  \definecolor{CYAN}{cmyk}{1,0,0,0}
  \definecolor{MAGENTA}{cmyk}{0,1,0,0}
  \definecolor{YELLOW}{cmyk}{0,0,1,0}
  }

\usepackage{bm}

\makeatother

\usepackage{babel}
\begin{document}

\title{Generating non-Rayleigh speckles with tailored intensity statistics}
\begin{abstract}
We experimentally generate speckle patterns with non-Rayleigh intensity
statistics using a phase-only spatial light modulator. By introducing
high order correlations to the input light fields we redistribute
the intensity among the speckle grains, while preserving the granular
structure of the pattern. Our method is versatile and allows for generating
speckle patterns with enhanced or diminished contrast in a controlled
manner. 
\end{abstract}

\author{Yaron Bromberg}

\email{yaron.bromberg@yale.edu}

\selectlanguage{english}%

\author{Hui Cao}

\email{hui.cao@yale.edu}

\selectlanguage{english}%

\affiliation{Department of Applied Physics, Yale University, New Haven, Connecticut
06520 USA}

\maketitle
Speckle patterns appear whenever a coherent wave impinges upon a scattering
sample. The granular structure of speckles results from the sensitivity
of the interference pattern to the relative phases of the scattered
partial waves. As speckle formation is essentially a wave phenomena,
it has been observed for a wide range of waves of different nature,
including ultrasonic waves \cite{US83}, microwaves \cite{GenackTransport,SAR_microwave81},
optical waves \cite{GoodmanBook}, X-rays \cite{Xray}, and matter
waves \cite{AtomSpeckles}. In spite of the diverse settings in which
speckles appear, they usually show universal statistical properties.
The apparent random distribution of the intensity mostly follows Rayleigh
statistics, i.e. the probability density function of the intensity
is a negative exponential \cite{GoodmanBook}. An interesting question
is whether the spatial structure of speckle patterns necessarily dictates
Rayleigh statistics, or perhaps it is possible to tailor the distribution
of the intensities among the speckle grains while maintaining the
random granular pattern? 

The reason that speckle patterns typically exhibit Rayleigh statistics is that Rayleigh statistics emerge under rather general conditions: the field is a sum of a large number of partial waves with independently varying amplitudes and phases, and the phases are uniformly distributed over a range of $2\pi$. In the weak scattering regime the latter condition is not satisfied, hence non-Rayleigh speckles with a low contrast and a strong DC background are formed \cite{Welford80}. Similarly, in the near-field zone of a scattering media where just a small number of scattered partial waves is detected and the relative phase of these waves does not cover the full  $2\pi$ range, low contrast speckles with nonuniversal statistics are observed \cite{Carminati95_nearfield,Dogariu03_NFcorrelations,DogariuPRE05}. In the strong scattering regime where the phases are uniformly distributed, deviations from Rayleigh statistics can be observed for a small number of partial waves, but the statistics approaches Rayleigh statistics as the number of partial waves is increased \cite{GoodmanAPO08}. For generating speckles with robust non-Rayleigh statistics that result from redistribution of the intensity among the speckle grains, we need to consider the interference of a large number of partial waves whose phases are uniformly distributed over a range of $2\pi$. In this case the only way to observe non-Rayleigh speckles is to make the complex amplitudes of the partial waves statistically dependent. Multiple scattering
can introduce mesoscopic correlations that modify Rayleigh statistics
due to strong fluctuations in the total power that is transmitted
through the sample \cite{vanRossum95}. However, the intensity distribution
for a single random configuration, i.e. the distribution of the intensities
between the speckle grains, still follows Rayleigh statistics \cite{Genack05,Muskens13}.
Similarly, non-Rayleigh statistics are observed when the total power
that is incident on a scattering sample fluctuates \cite{GoodmanBook}
or when two identical speckle patterns with a fluctuating relative
phase are interfered \cite{Zhang12_SubperBunched}, yet per speckle
realization the intensity is Rayleigh distributed. We on the other
hand are seeking to redistribute the intensities among the speckle
grains, so that each individual pattern will show non-Rayleigh statistics
and not just the ensemble average. 

In this Letter we show how to tailor the speckle statistics using
a phase-only spatial light modulator (SLM) that is illuminated by
a laser beam. The SLM pixels mimic scattering from a rough surface,
and the diffraction from each pixel corresponds to a partial wave
that is scattered from the SLM plane. We record the speckle patterns
at the Fourier plane of the SLM, where the intensity statistics are
determined by the statistical properties of the phase matrices that
are applied to the SLM. We developed a simple method for finding the
phase matrices that yield non-Rayleigh speckles, which we use for
tailoring the intensity distribution between the speckle grains. Since
speckle patterns are a valuable resource for both fundamental research
\cite{GenackTransport,MaretUCF,Weitz,Lugiato05_GI,SegevAL,Zimmermann99,Muskens13,Genack05}
and numerous applications \cite{Boas01,Gigan_SpeckleIllum,GiganContrast,LSIScheffold,Mertz08_HiLo,ScheffoldOEX11,Somekh04,Speckle_SI,Redding_Onchip},
a control of the intensity statistics can have a dramatic impact on
the way we use and analyze speckles. For example one can utilize tailored
speckles to synthesize the statistics of disordered optical potentials
for cold atoms and colloidal particles \cite{AspectAL,GiganTrapping,Dogariu_colloids},
or optimize the intensity statistics per application in speckle illumination
imaging \cite{Gigan_SpeckleIllum,Mertz08_HiLo,Somekh04,Speckle_SI}.
As speckle patterns are intimately related to the statistical properties
of thermal light sources, temporally fluctuating non-Rayleigh speckles
may also be regarded as a pseudothermal light source. Such a source
violates the Siegert relation and exhibits a bunching factor $g^{(2)}(0)\neq$2,
which enables exploring classical models of extra-bunching \cite{FabreExtraBunching,SuperBunchLeuchs}
and high-order ghost imaging \cite{BoydGI}.

\begin{figure}
\includegraphics[bb=-22bp -1bp 496bp 339bp,clip,width=1\columnwidth]{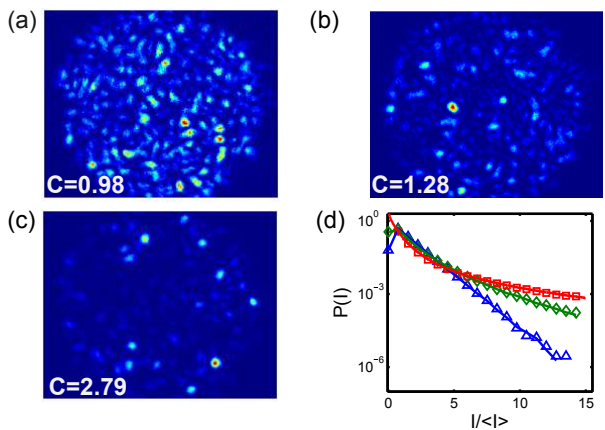}

\caption{Experimental generation of super-Rayleigh speckles. (a) - (c) Images
of the speckle patterns at the Fourier plane of the SLM, and (d) the
corresponding intensity distribution function. (a) Standard Rayleigh
speckles with a contrast of C=0.98 and a negative exponential intensity
distribution ((d), blue triangles). (b) Super-Rayleigh speckles with
a contrast of C=1.28, and an intensity distribution that decays slower
than the negative exponential ((d), green diamonds). The light is
concentrated at a fewer speckles grains compared to (a), yielding
non-Rayleigh statistics. (c) Higher contrast super-Rayleigh speckles
with C=2.79 and a long tailed intensity histogram ((d), red squares).
Solid lines are histograms obtained for the normalized speckle patterns
(see text). }
\end{figure}

To experimentally demonstrate our method for generating non-Rayleigh
speckles, we start with an example for enhancing the contrast of a
speckle pattern. We use a phase-only reflective SLM, which is illuminated
by a linearly polarized laser beam with diameter D=5mm. The SLM pixels
are grouped to macro pixels providing a control over 3000 independent
phase elements. We place the SLM at the front focal plane of a lens
and record the intensity pattern at the Fourier plane of the SLM by
imaging the back focal plane of the lens (see \cite{SI} for additional
details). When we apply to the SLM a random uncorrelated phase matrix
a Rayleigh speckle pattern is observed at the Fourier plane (figure
1a). To generate a speckle pattern with an enhanced contrast, we must
send to the SLM a phase matrix with correlated pixels. To find such
a matrix, we first numerically generate a high contrast speckle, for
example by squaring the field of a standard Rayleigh speckle, $E_{Ray}^{2}(x,y)$.
Next, we compute the inverse Fourier transform, and apply the phase
of the inverse Fourier transform to the SLM. Figure 1b shows the resulting
speckle pattern of a phase matrix computed in this way. It is clearly
seen that the distribution of the intensities among the speckle grains
is different than for Rayleigh speckles (figure 1a), a few grains
are much brighter than the rest. Indeed the intensity histogram collected
from a 1000 speckle realizations decays slower than a negative exponential,
featuring the high probability to obtain bright speckle grains (figure
1d, green diamonds). The contrast of the patterns, defined as $C=\sqrt{\left\langle I^{2}\right\rangle /\left\langle I\right\rangle ^{2}-1}$
where $\left\langle \cdots\right\rangle $ denotes spatial and ensemble
averaging, is C=1.28, significantly higher than the contrast measured
for the Rayleigh speckles C=0.98. We coin this kind of high contrast
patterns \textit{super-Rayleigh} speckles.

Instead of squaring the Rayleigh speckle field $E_{Ray}$ as in the
example above, in principle we can use any nonlinear transformation
$h(E_{Ray})$ to obtain non-Rayleigh speckle. Thus we can generate
super-Rayleigh speckles with a higher contrast by using, for example,
the fourth power of the speckle field $h(E_{Ray})=E_{Ray}^{4}$. We
send to the SLM the phase of the inverse Fourier transform of $E_{Ray}^{4}$,
and observe patterns with C=2.79 (figure 1c) and a long tailed intensity
histogram (figure 1d, red squares). We note that the long tailed intensity
histograms of super-Rayleigh speckles result from redistribution of
the intensity among the speckle grains and not because of fluctuations
in the total intensity of different speckle realizations. To verify
this, we normalize all the speckle patterns to have the same total
intensity, and show that the intensity histograms of the normalized
and non-normalized patterns are identical (figure 1d, solid lines). 

\begin{figure}
\includegraphics[bb=-20bp -1bp 339bp 118bp,clip,width=1\columnwidth]{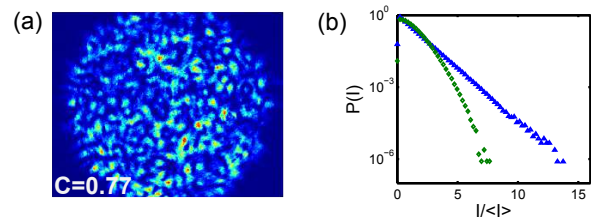}\label{fig:2}

\caption{(a) A speckle pattern measured at the Fourier plane of the SLM, using
a phase matrix that was designed to generate sub-Rayleigh speckles.
The low contrast of the pattern results from a more homogenous distribution
of the intensity among the speckle grains compared to Rayleigh speckles.
(b) Measured intensity distribution function of the sub-Rayleigh speckles
(green diamonds), which decays much faster than the negative exponential
observed for Rayleigh speckles (blue triangles). }
\end{figure}
The super-Rayleigh speckle statistics arises from concentrating the
light to a fewer bright grains compared to Rayleigh speckles. It is
interesting to explore the opposite regime which we coin \textit{sub-Rayleigh}
speckles, where the light is distributed in a more homogenous manner
among the grains. Intuitively, saturation of the intensity can reduce
the intensity fluctuations and the speckle contrast, while preserving
the granular structure of the speckles. Thus we use a nonlinear transformation
that saturates the amplitude of a Rayleigh speckle, but keeps its
phase untouched, $h(E_{Ray})=\sqrt{1-e^{-|E_{Ray}|^{2}}}e^{i\theta_{Ray}}$
where $\theta_{Ray}=arg(E_{Ray})$ is the phase of the Rayleigh speckle
field. Unlike super-Rayleigh speckles, where we applied to the SLM the phase of the inverse Fourier transform of $h(E_{Ray})$ and observed a pattern that to a good approximation matched $h(E_{Ray})$, for sub-Rayleigh speckles when we keep only the phase of the inverse Fourier transform of $h(E_{Ray})$ and disregard the  amplitude variation, we observe nearly standard Rayleigh speckles. This is because the amplitude modulation encodes much information of the transformed speckle. To transfer this information to phase modulation, we apply an iterative procedure based on the Gershberg-Saxton algorithm
\cite{GS}, where using Fourier transforms we numerically propagate
the field back and forth between the SLM plane and the Fourier plane.
At each iteration step we fix the amplitude at the Fourier plane to
be $\sqrt{1-e^{-|E_{Ray}|^{2}}}$, and we set the amplitude at the
SLM plane to match the amplitude of the beam profile that impinges
on the SLM. After 50 iterations the algorithm converges to a speckle
pattern at the Fourier plane with an intensity pattern that is proportional
to $1-e^{-|E_{Ray}|^{2}}$ and a corresponding phase matrix at the
SLM plane. We repeat the algorithm with different initial Rayleigh
speckles $E_{Ray}$ to generate a set of a phase matrices, which we
then send to the SLM. The recorded patterns indeed show a low contrast
(C=0.77), and the intensity histogram decays faster than a negative
exponential (figure 2). By comparing figure 2a to figure 1a, it can
be seen that intensity distribution among the speckle grains of the
sub-Rayleigh speckles is more homogenous, which is the origin of the
lower contrast. 

One interesting feature of the non-Rayleigh speckles is that the speckle
statistics changes as the beam propagates. All the results presented
so far were measured at the Fourier plane of the SLM, however, when
we scan the image plane away from the Fourier plane, the speckles
gradually return to Rayleigh statistics \cite{SI}. Figure 3 shows
the contrast of the recorded speckle patterns as a function of the
distance between the Fourier plane and the plane that is imaged by
the camera, showing that the contrast of the super-Rayleigh and sub-Rayleigh
speckles is axially dependent, whereas the contrast of the Rayleigh
speckle remains constant. The range over which the non-Rayleigh statistics are observed corresponds to the longitudinal length of a single speckle. This longitudinal length is equal to the Rayleigh range of an input Gaussian beam that is focused to a spot of the size of a single speckle \cite{Gatti08}. It is therefore determined by the envelope of the beam that illuminates the SLM, and it is not modified by phase-only modulation.

\begin{figure}
\label{fig:3}\includegraphics[bb=-10bp -1bp 490bp 167bp,clip,width=1\columnwidth]{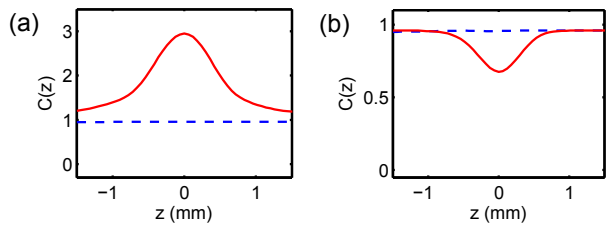}

\caption{Axially dependent speckle contrast. (a) The measured contrast of the
speckle patterns versus the distance from the Fourier plane, for Rayleigh
(blue dashed line), and super-Rayleigh (red line) speckles. (b) Same
as (a) for sub-Rayleigh speckles. The width of the contrast peak or
dip corresponds to the longitudinal length of a single speckle.}
\end{figure}
Observing non-Rayleigh statistics necessarily implies that the fields
at the SLM plane break at least one of the two conditions for observing
Rayleigh speckles: either the phases are not uniformly distributed
over $2\pi$, or the fields at different pixels are correlated. Figure
4a shows the histograms of the phases that were used to generate Rayleigh
speckles (blue triangles), super-Rayleigh speckles (green diamonds)
and sub-Rayleigh speckles (red squares). All three histograms are
constant over $[0,2\pi]$, which means that the non-Rayleigh statistics
must originate from correlations between the SLM pixels. We therefore
look at the correlation of the field at the SLM plane $G_{SLM}^{(1)}(\Delta_{x},\Delta_{y})=\left\langle E_{SLM}(x,y)E_{SLM}^{*}(x+\Delta_{x},y+\Delta_{y})\right\rangle $.
We obtain $G_{SLM}^{(1)}$ using the Wiener-Khinchin theorem \cite{MandelWolf},
by computing the Fourier transform of the averaged intensity pattern
at the Fourier plane of the SLM. The $G_{SLM}^{(1)}$ curves for super-
and sub-Rayleigh speckles are presented in figure 4b, showing that
fields at different macropixels are uncorrelated. The question is
then how can uncorrelated fields with phases that are uniformly distributed
over $2\pi$ yield non-Rayleigh statistics? The answer is that $G_{SLM}^{(1)}(\Delta)=0$
indicates that fields separated by a distance $\Delta$ are uncorrelated,
but it does not necessarily mean that the fields are statistically
independent. For non-Gaussian random fields, higher order correlations
can exist even if the first order correlation vanishes \cite{MandelWolf}. 

To investigate the role of higher order correlations in the formation
of non-Rayleigh speckles, we study numerically the contribution of
the second-order field correlations to the contrast of the speckle
patterns. Since the laser beam is linearly polarized and the scattering
angles from the SLM are too small to introduce radial polarization,
we consider scalar fields. A partial wave that diffracts from an SLM
pixel at position \textbf{$\mathbf{r}$} to position \textbf{${\bm{\rho}}$}
in the Fourier plane is proportional to $e^{i(\psi_{\mathbf{r}}-2\pi\ensuremath{{\bm{\rho}}}\mathbf{r/\lambda}f\mathbf{)}}$,
where $\psi_{\mathbf{r}}$ represents the SLM phase at \textbf{r}.
The square of the contrast of the speckle pattern at the Fourier plane is therefore:
\begin{eqnarray}
C^2 & \equiv & \langle I^{2}\rangle-\langle I\rangle^{2}=\left\langle \int d^{2}\ensuremath{{\bm{\rho}}}|E(\mathbf{\ensuremath{{\bm{\rho}}}})|^{4}\right\rangle _{e}-1\label{eq:1}\\
 & = & \frac{1}{N^{2}}\sum_{\mathbf{r_{1},}\mathbf{r_{2}},\mathbf{r_{3}}\mathbf{r_{4}}}\mathbf{\left\langle e^{i(\psi_{\mathbf{r1}}+\psi_{r2}-\psi_{r3}-\psi_{r4}}\right\rangle _{e}\delta_{r_{1}+r_{2}-r_{3},r_{4}}\,}-1\nonumber 
\end{eqnarray}
where $N$ is the number of SLM pixels, $\left\langle \cdots\right\rangle _{e}$
denotes ensemble averaging, and $\left\langle I\right\rangle $ is
normalized to 1. The sum on the right hand side is in fact a sum over
the second order correlation between the fields at all the SLM pixels:
$\sum G_{SLM}^{(2)}(\mathbf{r_{1}},\mathbf{r_{2}},\mathbf{r_{3}},\mathbf{\mathbf{r_{1}}+\mathbf{r_{2}}-\mathbf{r_{3}}})$.
We can decompose this sum into 4 terms, $C^2=\Gamma_{1}^{(2)}+\Gamma_{2}^{(2)}+\Gamma_{3}^{(2)}+\Gamma_{4}^{(2)}-1$,
where $\Gamma_{p}^{(2)}$ is the second order correlation between
fields at $p$ \textit{different} pixels, e.g. $\Gamma_{4}^{(2)}=\sum_{\mathbf{r_{1}}\neq\mathbf{r_{2}}\neq\mathbf{r_{3}}\neq\mathbf{r_{4}}}G_{SLM}^{(2)}(\mathbf{r_{1}},\mathbf{r_{2}},\mathbf{r_{3}},\mathbf{r_{4}=\mathbf{r_{1}}+\mathbf{r_{2}}-\mathbf{r_{3}}})$.
For Rayleigh speckles that are formed by statistically independent
pixels, only correlations of fields at the same pixel contribute to
the sum, hence $\Gamma_{1}^{(2)}=1/N$, $\Gamma_{2}^{(2)}=2-2/N$
and $\Gamma_{3}^{(2)}=\Gamma_{4}^{(2)}=0$. For non-Rayleigh speckles,
we evaluate $\Gamma_{p}^{(2)}$ numerically by calculating the second
order correlations $G_{SLM}^{(2)}(\mathbf{r_{1}},\mathbf{r_{2}},\mathbf{r_{3}},\mathbf{r_{4}=\mathbf{r_{1}}+\mathbf{r_{2}}-\mathbf{r_{3}}})$
for all the combinations of the indices $\{\mathbf{r_{1},r_{2},r_{3}},\mathbf{r_{4}}\}$,
and decomposing the results into the four terms $\Gamma_{p}^{(2)}$
according to the number of different indices that appear in each combination.
The numerical results for 5000 phase patterns with $100$ pixels are
summarized in Table 1. It is seen that the source of the non-Rayleigh
statistics is $\Gamma_{4}^{(2)},$ i.e. the cross-correlation between
four \textit{different} pixels. 

\begin{figure}
\label{fig:4}\includegraphics[width=1\columnwidth]{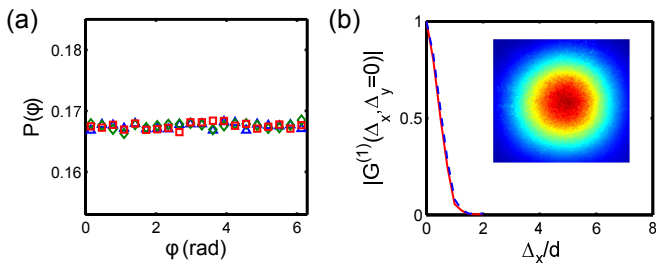}\caption{(a) Histograms of the phases used to generate Rayleigh (blue triangles),
super-Rayleigh (green diamonds) and sub-Rayleigh (red squares) speckles.
All three histograms are constant over $2\pi$. (b) The field correlation
at the SLM plane $G_{SLM}^{(1)}$ for super-Rayleigh speckle (blue
dashed line) and sub-Rayleigh speckle (red solid line). $d=80\mu m$
is the size of one macro pixel. Fields separated by more than one
macropixel are uncorrelated, indicating that the source of the non-Rayleigh
statistics is in higher order correlations. The inset shows the average
intensity for the sub-Rayleigh speckles, from which $G_{SLM}^{(1)}$
was computed. }
\end{figure}

In Eq. (1) the high order correlations of the field at the SLM plane
are linked to the intensity statistics at the Fourier plane. However,
for points that are not in the Fourier plane an additional quadratic
phase factor $e^{i\pi z(\mathbf{r_{1}^{2}}+\mathbf{r_{2}^{2}}-\mathbf{r_{3}^{2}}-\mathbf{r_{4}^{2}})/\lambda f^{2}}$
multiplies each term in the right hand side of Eq. (1), where $z$
is axial the distance from the Fourier plane. Due to this quadratic phase the high order correlations do not add in phase and their contribution to the intensity statistics vanishes. This is the reason why we observe non-Rayleigh statistics only at the Fourier plane of the SLM. We note that by adding a quadratic phase to the SLM it is possible to shift the axial position of the Fourier plane where the speckle statistics is non-Rayleigh.

\begin{center}
\begin{table}
\begin{centering}
\begin{tabular}{lcccc}
\hline 
\multirow{1}{*}{\begin{sideways}
\end{sideways}} & $\Gamma_{1}^{(2)}$ & $\Gamma_{2}^{(2)}$ & $\Gamma_{3}^{(2)}$ & \textbf{$\Gamma_{4}^{(2)}$}\tabularnewline
\hline 
Rayleigh & 0.0 & 2.0 & 0.0 & \textbf{0.0 }\tabularnewline
Super-Rayleigh$h(E)=E^{2}$~~ & 0.0 & 2.0 & 0.0 & \textbf{0.7}\tabularnewline
Super-Rayleigh$h(E)=E^{4}$~ & 0.0 & 2.0 & 0.0 & \textbf{5.6}\tabularnewline
Sub-Rayleigh & 0.0 & 2.0 & 0.0 & \textbf{-0.4}\tabularnewline
\hline 
\end{tabular}
\par\end{centering}

\caption{Numerical evaluation of $\Gamma_{p}^{(2)},$ the second order correlation
between fields at $p$ different pixels, for speckles of different
statistics. The cross-correlation between fields at four different
pixels $\Gamma_{4}^{(2)}$ is the source of the non-Rayleigh statistics. }
\end{table}

\par\end{center}

In conclusion, we developed a method to generate speckle patterns
with controlled intensity statistics. In contrast to previous works
which used an SLM to control the spatial coherence of light fields
by shaping the spatial frequency distribution \cite{FractalSpeckles,Waller12,FleischerPRL12}
or synthesizing the coherent mode decomposition of the field \cite{Boyd13_SpatialCoherenceControl},
we use the SLM to redistribute the intensity among the speckle grains.
We link the observed non-Rayleigh statistics to higher order correlations
of the scattered partial waves, paving the way towards a better understanding
of the information carried by speckles that are scattered from disordered
samples with structural correlations. Non-Rayleigh speckles also offer
new opportunities for speckle illumination applications. On one hand,
the axially varying contrast of super-Rayleigh speckles can be utilized
for achieving better optical sectioning in speckle illumination microscopy.
On the other hand, the homogenous distribution of sub-Rayleigh speckles
are attractive for imaging modalities that utilize speckle illumination,
as it can reduce the number of projections required for averaging
the speckles to observe a smooth image and enhance the image acquisition
rate. Therefore, tailored speckles allows optimizing the intensity
statistics for target applications. 

We thank Eric Dufresne, Brandon Redding, Sebastien Popoff and Arthur
Goetschy for fruitful discussions. This work is supported partially
by NSF under the Grant Nos. ECCS1128542 and DMR1205307.

\end{document}

\title{Supplementary Material: Generating non-Rayleigh speckles with tailored
intensity statistics}

\author{Yaron Bromberg}

\email{yaron.bromberg@yale.edu}

\author{Hui Cao}

\email{hui.cao@yale.edu}

\affiliation{Department of Applied Physics, Yale University, New Haven, Connecticut
06520 USA}

\maketitle
\renewcommand{\figurename}{Figure S}

\section*{experimental setup}

A phase-only reflective SLM (Hamamatsu, LCOS-SLM) is illuminated by
a collimated linearly polarized laser beam (Coherent OBIS, $\lambda$=640nm),
which is expanded and clipped by an iris to uniformly cover a circular
area on the SLM with diameter D=5mm. As shown schematically in figure
S1, the SLM pixels are grouped to macropixels with a pitch of 80$\mu m$,
providing a control over 3000 independent phase elements. The SLM
is placed at the front focal plane of a lens $L_{1}$ ($f_{\ensuremath{1}}$=50cm),
and the back focal plane of the lens is re-imaged onto a 12bit CCD
camera (Manta, Allied Vision) using 2 lenses, $L_{2}$ and $L_{3}$,
in a 4f configuration ($f_{2}$=10cm, $f_{3}$=15cm). Axial scans
of the image plane was performed by adding to the beam path two identical
microscope objectives (x10, NA=0.25) and scanning the position of
one of the objectives using a motorized stage.

\begin{figure}[H]
\includegraphics[bb=-20bp -1bp 253bp 141bp,clip,width=1\columnwidth]{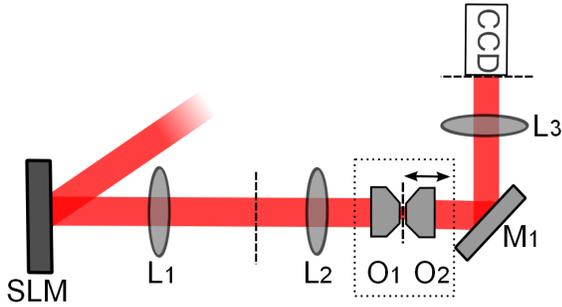}\label{fig:S1}

\caption{Schematic of the experimental setup. A phase-only SLM is placed at
the front focal plane of a lens ($L_{1})$. The Fourier transform
of the SLM plane is formed at the back focal plane of $L_{1}$, and
is reimaged onto a CCD camera using a pair of lens ($L_{2},$$L_{3}$)
in a 4f configuration. The axial variation of the speckle pattern
was recorded by inserting a pair of identical microscope objectives
($O_{1}$,$O_{2}$) after lens $L_{2}$, and scanning the axial position
of $O_{2}$ using a motorized stage. The dashed lines mark the Fourier
plane and its two conjugated planes. }
\end{figure}

\section*{Three dimensional pattern of non-Rayleigh speckles}

The non-Rayleigh speckles have a non-homogenous three-dimensional
structure, the deviation from Rayleigh statistics is observed only
in the vicinity of the Fourier plane of the SLM. Images of the patterns
measured at the Fourier plane, and 3mm away from the Fourier plane
are presented in Figure S2. While the patterns of the super-Rayleigh
and sub-Rayleigh speckles clearly exhibit non-Rayleigh statistics
at the Fourier plane, away from the Fourier plane they resemble standard
speckle patterns with Rayleigh statistics. Movies showing the the
transition from the non-Rayleigh to the Rayleigh statistics as the
speckle propagates are recorded by scanning the axial position of
the plane that is imaged onto the camera, and attached \cite{video}.

\begin{figure}[H]
\includegraphics[bb=-10bp -1bp 283bp 321bp,clip,width=1\columnwidth]{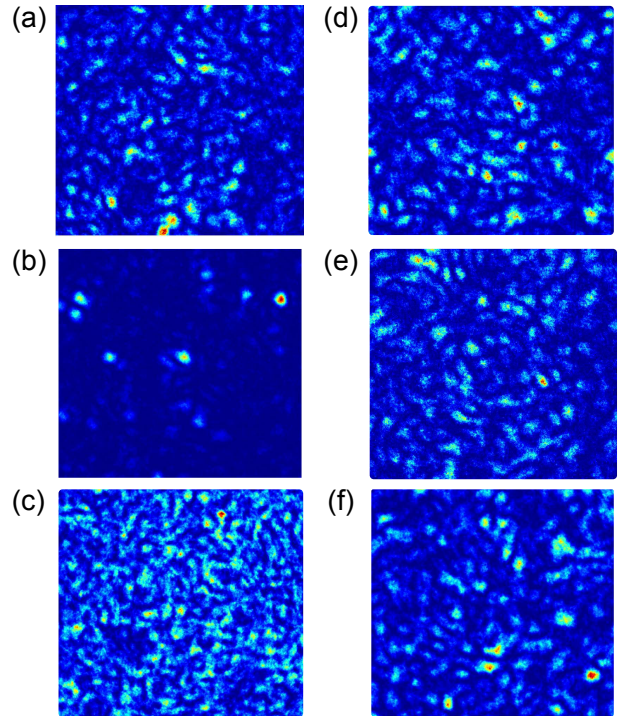}\label{fig:S2}

\caption{Images recorded at the Fourier plane (left column), and 3mm away from
the Fourier plane (right column), for Rayleigh and non-Rayleigh speckles.
Rayleigh speckles (a,b) show the same statistics in both planes, where
as super-Rayleigh (c,d) and sub-Rayleigh (e,f) speckles transform
into Rayleigh speckles outside of the Fourier plane.}
\end{figure}

\section*{Deviations from Rayleigh statistics of the local contrast}
In this section we study whether the non-Rayleigh speckle statistics that we measure are in fact locally Rayleigh statistics, but with a spatially dependent average intensity.  We have already shown in the main text that the ensemble averaged intensity of the non-Rayleigh speckles is spatially homogenous, resulting in a delta like field-field correlations for the field at the SLM plain (figure 4b). This however does not rule out the possibility that the statistics are locally Rayleigh statistics, and that the local averaged intensity changes for different realizations.  
In order to to check whether this is the case, we perform the following analysis of the experimental data. For each speckle realization, we choose a region of interest (ROI) that is centered around the brightest speckle in the frame. We calculate the local contrast inside the ROI by computing $C_r=\sqrt{\left\langle I^2 \right\rangle_s/\left\langle I \right\rangle_s^2-1}$, where $\left\langle\cdots\right\rangle_s$ denotes spatial averaging over the ROI area. Next we compute the ensemble average of $C_r$ over all the speckle realizations $C=\left\langle C_r \right\rangle_e$ . If the non-Rayleigh statistics originated from local Rayleigh statistics, the local contrast should have been equal to C=1 for some range of sizes of the ROI. In order to find the minimal size of the ROI which covers a sufficient number of speckle grains so that the measure of the local contrast would be statistically meaningful, we compute the local contrast also for the Rayleigh speckles (figure S3, blue dotted line). We find that when the ROI covers more than ~40 speckle grains the contrast of the Rayleigh speckle converges to C=1. For this size of the ROI it is clearly seen that the contrast of the super-Rayleigh (green dashed line) and sub-Rayleigh (red solid line) speckles are significantly different than C=1, i.e. even the local contrast shows non-Rayleigh statistics. 

\begin{figure}[H]
\includegraphics[width=1\columnwidth]{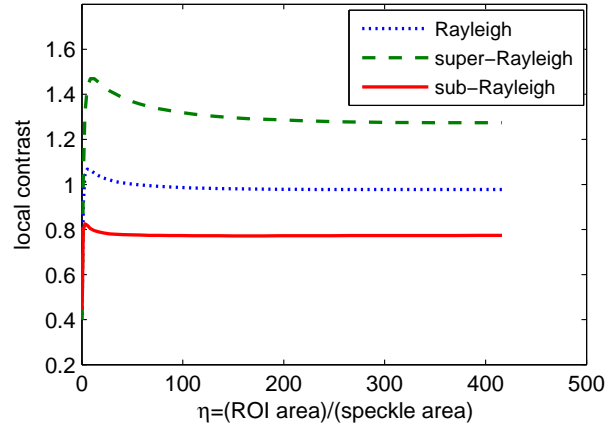}\label{fig:S3}

\caption{The local contrast as a function of number of speckle grains in the ROI (defined by the ratio of the ROI area and the speckle grain area). When the area of the ROI contains more than 40 speckle grains $(\eta>40)$, the contrast for Rayleigh speckles converges to 1, indicating that for $\eta>40$ the definition of the local contrast is statistically meaningful (blue dotted line). In this regime the local contrast of the super-Rayleigh (dashed green line) and sub-Rayleigh (solid red line) speckles exhibits non-Rayleigh statistics, proving that the non-Rayleigh statistics does not result from Rayleigh speckles with a local varying average intensity.}
\end{figure}